\documentclass[11pt]{book}

\usepackage{amssymb}
\usepackage{amsmath}
\usepackage{amsfonts}
\usepackage{epsfig}

\makeatletter

\renewenvironment{thebibliography}[1]
     {\section*{\bibname}%
      \@mkboth{\MakeUppercase\bibname}{\MakeUppercase\bibname}%
      \list{\@biblabel{\@arabic\c@enumiv}}%
           {\settowidth\labelwidth{\@biblabel{#1}}%
            \leftmargin\labelwidth
            \advance\leftmargin\labelsep
            \@openbib@code
            \usecounter{enumiv}%
            \let\p@enumiv\@empty
            \renewcommand\theenumiv{\@arabic\c@enumiv}}%
      \sloppy
      \clubpenalty4000
      \@clubpenalty \clubpenalty
      \widowpenalty4000%
      \sfcode`\.\@m}
     {\def\@noitemerr
       {\@latex@warning{Empty `thebibliography' environment}}%
      \endlist}

\makeatother

\newcommand{\sect}[1]{\setcounter{equation}{0}\section{#1}}

\renewcommand\bibname{References}

\DeclareMathOperator*{\Tr}{Tr}
\DeclareMathOperator{\Ai}{Ai}
\DeclareMathOperator{\Bes}{Bes}
\DeclareMathOperator{\PII}{PII}
\DeclareMathOperator{\Pear}{Pear}
\DeclareMathOperator*{\Pf}{Pf}
\DeclareMathOperator*{\sgn}{sgn}
\DeclareMathOperator*{\supp}{supp}
\DeclareMathOperator*{\diag}{diag}
\DeclareMathOperator*{\Repart}{Re}
\DeclareMathOperator*{\Impart}{Im}

\newcommand{\ds}{\displaystyle}

\begin{document}

\setlength{\baselineskip}{5.0mm}


\setcounter{chapter}{5}
\chapter[Universality]{Universality{\LARGE\footnote{To appear as Chapter 6 in: 
Oxford Handbook of Random Matrix Theory, (G. Akemann, J. Baik, and P. Di Francesco, eds.),
Oxford University Press, 2011.}}}
\thispagestyle{empty}

\noindent
{\sc A. B. J. Kuijlaars}\footnote{
Department of Mathematics,
Katholieke Universiteit Leuven, Celestijnenlaan 200B, 3001 Leuven, Belgium. 
The author is supported in part by FWO-Flanders projects G.0427.09 and
G.0641.11, by K.U. Leuven research grant OT/08/33, by the Belgian Interuniversity
Attraction Pole P06/02, and by a grant from the Ministry of Education and
Science of Spain, project code MTM2005-08648-C02-01.}

\begin{center}
{\bf Abstract}
\end{center}
Universality of eigenvalue spacings is one of the basic
characteristics of random matrices. We give the precise meaning of
universality and discuss the standard universality classes (sine, Airy, Bessel)
and their appearance
in unitary, orthogonal, and symplectic  ensembles.
The Riemann-Hilbert problem for orthogonal polynomials is
one possible tool to derive universality
in unitary random matrix ensembles. An overview is presented of the
Deift/Zhou steepest descent analysis of the Riemann-Hilbert problem
in the one-cut regular case.
Non-standard universality classes that arise at singular points
in the spectrum are discussed at the end.


\sect{Heuristic meaning of universality} \label{heuristic}

Universality of eigenvalue spacings is one of the basic characteristic
features of random matrix theory. It was the main motivation
of Wigner to introduce random matrix ensembles to model energy levels
in quantum systems.

On a local scale the eigenvalues of random matrices show a repulsion
which implies that it is highly unlikely that eigenvalues are very close
to each other. Also very large gaps between neighboring eigenvalues
are unlikely. This is in sharp contrast to points that are sampled
independently from a given distribution. Such points exhibit Poisson statistics
in contrast to what is now called GUE/GOE/GSE statistics.
These new random matrix theory type statistics have been observed
(and sometimes proved) in many other mathematical and physical systems
as well, see \cite{Deift2}, and are therefore universal in a sense that
goes beyond random matrix theory. Examples outside of random matrix
theory are discussed in many chapters of this handbook.
In this chapter the discussion of the universality is however restricted to
random matrix theory.

The universality conjecture in random matrix theory says that the
local eigenvalue statistics of many random matrix ensembles are the same,
that is, they do not depend on the exact probability distribution that is put
on a set of matrices, but only on some general characteristics of the ensemble. The universality
of local eigenvalue statistics is a phenomenon that takes place for
large random matrices. In a proper mathematical formulation it
is a statement about a certain limiting behavior as the size of the matrices
tends to infinity.

The characteristics that play a role in the determination of the
universality classes are the following.
\begin{itemize}
\item Invariance properties of the ensemble, of which the prototype
is invariance with respect to orthogonal, unitary, or unitary-symplectic conjugation.
As discussed in Chapters 4 and 5 of this handbook,
these lead to random matrix ensembles with an explicit joint eigenvalue density
of the form
\begin{equation} \label{eq:invariantpdf}
    p(x_1, \ldots, x_n) =
    \frac{1}{Z_{n,\beta}} \prod_{1 \leq i < j \leq n} |x_j-x_i|^{\beta} \prod_{j=1}^n e^{- V(x_j)}
    \end{equation}
where $\beta=1,2,4$ corresponds to orthogonal, unitary,  and
symplectic ensembles, respectively. The case $V(x) = \frac{1}{2}
x^2$ gives the Gaussian ensembles GOE, GUE, and GSE. The local
eigenvalue repulsion increases with $\beta$ and different values
of $\beta$ give rise to different universality results.

\item An even more basic characteristic is the distinction between
random matrix ensembles of real symmetric or complex Hermitian
matrices. Even without the invariance assumption, the universality
of local eigenvalue statistics is conjectured to hold. This has
been proved recently for large classes of  Wigner
ensembles,  i.e., random matrix ensembles with independent,
identically distributed entries, see \cite{EPRSY,ERSTVY,TaoVu} and
the survey \cite{Erd}. See Chapter 21 of this handbook for
more details.  Universality is also
expected to exist in classes of non-Hermitian matrices, see Chapter 18.

\item Another main characteristic is the nature of the point around which
the local eigenvalue statistics are considered. A typical point in the
spectrum is such that, possibly after appropriate scaling, the limiting
mean eigenvalue density is positive. Such a point is in the bulk of
the spectrum, and universality around such a point is referred to
as bulk universality.

At edge points of the limiting spectrum the limiting
mean eigenvalue density typically vanishes like a square root, and at
such points a different type of universality is expected, which is known as
(soft) edge universality. At natural edges of the spectrum, such as
the point zero for ensembles of positive definite matrices, the limiting
mean eigenvalue density typically blows up like the inverse of a square root,
and universality at such a point is known as hard edge universality.

\item
At very special points in the limiting spectrum
the limiting mean eigenvalue density may exhibit singular behavior. For example,
the density may vanish at an isolated point in the interior of
the limiting spectrum, or it may vanish to higher order than square root
at soft edge points. This non-generic behavior
may take place in ensembles of the form \eqref{eq:invariantpdf}
with a varying potential $NV$
\begin{equation} \label{eq:invariantpdf2}
    p(x_1, \ldots, x_n) =
    \frac{1}{Z_{n,\beta}} \prod_{1 \leq i < j \leq n} |x_j-x_i|^{\beta} \prod_{j=1}^n e^{- N V(x_j)}
    \end{equation}
as $n, N \to \infty$ with $n/N \to 1$.

At such special points one expects other universality classes
determined by the nature of the vanishing of the limiting mean
eigenvalue density at that point, see also a discussion in Chapter 14.
\end{itemize}

In the rest of this chapter we first give the precise meaning of the notion
of universality and we discuss the limiting kernels (sine, Airy and Bessel)
associated with the bulk universality and the edge universality for the
unitary, orthogonal, and symplectic universality classes.
In Section \ref{sect:unitary} we discuss unitary matrix ensembles
in more detail and we show that universality in these ensembles
comes down to the convergence of the properly
scaled eigenvalue correlation kernels. In Section \ref{sect:RHmethod}
we discuss the Riemann-Hilbert method in some detail. The Riemann-Hilbert
method is one of the main methods to prove universality in unitary
ensembles. In the final Section \ref{sect:non-standard} we discuss
certain non-standard universality classes that arise at singular
points in the limiting spectrum. We describe the limiting
kernels for each of the  three types
of singular points, namely interior singular points, singular
edge points, and exterior singular points.

Other approaches to universality are detailed in Chapter 21 for Wigner
ensembles and in chapter 16 using loop equation techniques.

\sect{Precise statement of universality} \label{sect:precise}

For a probability density $p_n(x_1, \ldots, x_n)$ on $\mathbb R^n$
that is symmetric (i.e., invariant under permutations of coordinates),
let us define the $k$-point correlation function by
\begin{equation} \label{eq:kpointcorrelation}
    R_{n,k}(x_1, \ldots, x_k) = \frac{n!}{(n-k)!} \int \cdots \int p_n(x_1, \ldots, x_n)
    \, dx_{k+1} \cdots dx_n.
    \end{equation}
Up to the factor $n!/(n-k)!$ this is the $k$th marginal distribution of $p_n$.

Fix a reference point $x^*$ and a constant $c_n > 0$. We center the points
around $x^*$ and scale by a factor $c_n$, so that $(x_1, \ldots, x_n)$
is mapped to
\[ (c_n(x_1 - x^*), \ldots, c_n (x_n - x^*)). \]
These centered and scaled points have the following
rescaled $k$-point correlation functions
\begin{equation} \label{eq:kpointrescaled}
    \frac{1}{c_n^k}
    R_{n,k}\left(x^* + \frac{x_1}{c_n}, x^* + \frac{x_2}{c_n}, \ldots, x^* + \frac{x_k}{c_n} \right).
\end{equation}

The universality is a property of a sequence $(p_n)$ of symmetric probability
density functions. Universality at $x^*$ means that
for a suitably chosen sequence $(c_n)$ the rescaled $k$-point correlation functions
\eqref{eq:kpointrescaled} have a specific limit as $n \to \infty$.
The precise limit determines the universality class.

As discussed in  chapter 4 for determinantal point processes,
the main spectral statistics such as gap probabilities and eigenvalue spacings
can be expressed in terms of the $k$-point correlation functions.
If the limits \eqref{eq:kpointrescaled} exist and belong to a certain
universality class, then this also leads to the universal behavior
of the local eigenvalue statistics.

\subsection{Unitary universality classes}
The unitary universality classes are characterized by the fact that
the limit of \eqref{eq:kpointrescaled} can be expressed as a $k \times k$
determinant
\[ \det \left[ K(x_i,x_j) \right]_{1 \leq i,j \leq k} \]
involving a kernel $K(x,y)$, which is called the eigenvalue correlation kernel.

\paragraph{Bulk universality}
A sequence of symmetric probability density functions $(p_n)$ then has bulk universality
at a point $x^*$ if there exists a sequence $(c_n)$,
so that for every $k$, the limit of \eqref{eq:kpointrescaled} is given
by
\[ \det \left[ K^{\sin}(x_i, x_j) \right]_{1 \leq i, j \leq k} \]
where
\begin{equation} \label{eq:sinekernel}
    K^{\sin}(x,y) = \frac{\sin \pi(x-y)}{\pi(x-y)}
    \end{equation}
is the sine kernel. Bulk universality is also known as GUE statistics.

\paragraph{Edge universality}
The (soft) edge universality holds at $x^*$ if the limit of \eqref{eq:kpointrescaled}
is equal to
\[ \det \left[ K^{\Ai}(x_i, x_j) \right]_{1 \leq i, j\leq k} \]
where
\begin{equation} \label{eq:Airykernel}
    K^{\Ai}(x,y) =  \frac{\Ai(x) \Ai'(y) - \Ai'(x) \Ai(y)}{x-y}
    \end{equation}
is the Airy kernel. Here $\Ai$ denotes the Airy function, which is
the solution of the Airy differential equation $y''(x) = xy(x)$
with asymptotics
\[ \Ai(x) = \frac{1}{2\sqrt{\pi}x^{1/4}}  e^{- \frac{2}{3} x^{3/2}} \left( 1 + O(x^{-3/2})\right),
    \qquad \text{as } x \to + \infty.
\]
The Airy kernel is intimately related to the Tracy-Widom distribution
for the largest eigenvalue in random matrix theory \cite{TW1}.

\paragraph{Hard edge universality}
A hard edge is a boundary for the eigenvalues that is part of the model.
For example, if the random matrices are real symmetric (or complex Hermitian) positive definite then
all eigenvalues are non-negative and zero is a hard edge.

The hard edge universality holds at $x^*$ if the limit of \eqref{eq:kpointrescaled}
is equal to
\[ \det \left[ K^{\Bes, \alpha}(x_i, x_j) \right]_{1 \leq i, j\leq k} \]
where
\begin{equation} \label{eq:Besselkernel}
    K^{\Bes,\alpha}(x,y) =
    \frac{J_{\alpha}(\sqrt{x}) \sqrt{y} J_{\alpha}'(\sqrt{y}) -
    \sqrt{x} J_{\alpha}'(\sqrt{x}) J_{\alpha}(\sqrt{y})}{2(x-y)},
    \qquad x, y > 0,
    \end{equation}
and $J_{\alpha}$ is the usual Bessel function of order $\alpha$.

The Bessel kernels depend on the parameter $\alpha > -1$ which may
be interpreted as a measure of the interaction with the hard edge.
The bigger $\alpha$, the more repulsion from the hard edge.

\subsection{Orthogonal and symplectic universality classes}

The orthogonal and symplectic universality classes are characterized by the fact that
the limit of \eqref{eq:kpointrescaled} is expressed as a Pfaffian
\begin{equation} \label{eq:Pfaffianlimit}
    \Pf \left[ K(x_i,x_j) \right]_{1 \leq i, j \leq k}
    \end{equation}
of a $2 \times 2$ matrix kernel
\begin{equation} \label{eq:matrixkernel}
    K(x,y) = \begin{pmatrix} K_{11}(x,y) & K_{12}(x,y) \\
    K_{21}(x,y) & K_{22}(x,y) \end{pmatrix}.
    \end{equation}
Recall that the Pfaffian $\Pf(A)$ of a $2k \times 2k$ skew symmetric matrix $A$
is such that
\[ \Pf(A) = \sqrt{\det(A)}. \]
The matrix in  \eqref{eq:Pfaffianlimit} is a $2k \times 2k$ matrix written
as a $k \times k$ block matrix with $2 \times 2$ blocks.
It is skew symmetric provided that $K(y,x) = - K(x,y)^T$
where $^T$ denotes the matrix transpose.

\paragraph{Bulk universality}
A sequence of symmetric probability density functions $(p_n)$ has
orthogonal/symplectic bulk universality
at a point $x^*$ if there exists a sequence $(c_n)$,
so that for every $k$, the limit of \eqref{eq:kpointrescaled} is given
by
\[ \Pf \left[ K^{\sin, \beta}(x_i, x_j) \right]_{1 \leq i, j \leq k} \]
with
\[ K^{\sin,\beta=1}(x,y) =
    \begin{pmatrix}
    \ds - \frac{\partial}{\partial x} \frac{\sin \pi(x-y)}{\pi(x-y)}
    & \ds \frac{\sin \pi(x-y)}{\pi(x-y)} \\[10pt]
    \ds - \frac{\sin \pi(x-y)}{\pi(x-y)} &
    \ds\int_0^{x-y} \frac{\sin \pi t}{\pi t} dt - \frac{1}{2} \sgn(x-y)
    \end{pmatrix} \]
where
$\sgn(t) = 1,0, -1$ depending on whether $t > 0$, $t=0$, or $t< 0$,
in the case of orthogonal (i.e., $\beta=1$) bulk universality, and
\[ K^{\sin, \beta=4}(x,y) =
    \begin{pmatrix} \ds -\frac{\partial}{\partial x} \frac{\sin 2\pi(x-y)}{2\pi(x-y)}
    & \ds \frac{\sin 2\pi(x-y)}{2\pi(x-y)} \\[10pt]
    - \ds \frac{\sin 2\pi(x-y)}{2\pi(x-y)} &
    \ds \int_0^{x-y} \frac{\sin 2\pi t}{2\pi t} dt
    \end{pmatrix} \]
in the case of symplectic (i.e., $\beta = 4$) bulk universality.

\paragraph{Edge universality}
The orthogonal/symplectic (soft) edge universality holds at $x^*$
if the limit of \eqref{eq:kpointrescaled}
is equal to
\[ \Pf \left[ K^{\Ai, \beta}(x_i,x_j) \right]_{1 \leq i, j \leq k} \]
with
\begin{equation} \label{eq:Airykernel1}
\begin{aligned}
    K_{11}^{\Ai,\beta=1}(x,y) & = \frac{\partial }{\partial y} K^{\Ai}(x,y) +  \frac{1}{2} \Ai(x) \Ai(y) \\
    K_{12}^{\Ai,\beta=1}(x,y) & = K^{\Ai}(x,y) + \frac{1}{2} \Ai(x) \cdot \int_{-\infty}^y \Ai(t) \, dt \\
    K_{21}^{\Ai,\beta=1}(x,y) & = - K_{12}^{\Ai,\beta=1}(x,y) \\
    K_{22}^{\Ai, \beta=1}(x,y) & = - \int_x^{\infty} K^{\Ai}(t,y) \, dt
        - \frac{1}{2} \int_x^y \Ai(t) \, dt \\
        & \qquad + \frac{1}{2} \int_x^{\infty} \Ai(t) \, dt
            \cdot \int_y^{\infty} \Ai(t) \, dt - \frac{1}{2} \sgn(x-y)
            \end{aligned}
            \end{equation}
in the case of orthogonal edge universality, and
\begin{equation} \label{eq:Airykernel4}
\begin{aligned}
    K_{11}^{\Ai,\beta=4}(x,y) & = \frac{1}{2} \frac{\partial }{\partial y} K^{\Ai}(x,y) +  \frac{1}{4} \Ai(x) \Ai(y) \\
    K_{12}^{\Ai,\beta=4}(x,y) & = \frac{1}{2} K^{\Ai}(x,y) - \frac{1}{4} \Ai(x) \cdot \int_y^{\infty} \Ai(t) \, dt \\
    K_{21}^{\Ai,\beta=4}(x,y) & = - K_{12}^{\Ai,\beta=4}(x,y) \\
    K_{22}^{\Ai, \beta=4}(x,y) & = - \frac{1}{2} \int_x^{\infty} K^{\Ai}(t,y) \, dt
        + \frac{1}{4} \int_x^{\infty} \Ai(t) \, dt
            \cdot \int_y^{\infty} \Ai(t) \, dt
            \end{aligned}
        \end{equation}
in the case of symplectic edge universality.
The kernel $K^{\Ai}$ that appears in \eqref{eq:Airykernel1} and \eqref{eq:Airykernel4}
is the Airy kernel from \eqref{eq:Airykernel}.

\paragraph{Hard edge universality}

The orthogonal/symplectic hard edge universality is expressed in terms
of $2 \times 2$ kernels with Bessel functions. See Forrester \cite{For2}
for the precise statement.

\subsection{Determinantal and Pfaffian point processes}

The universality limits have the characteristic properties of determinantal
or Pfaffian point processes, see \cite{For2,Sos} and Chapter 11 of this handbook.

If the probability densities $p_n$ on $\mathbb R^n$ also arise
from determinantal point processes, then the statement of universality
can be expressed as the convergence of the corresponding correlation
kernels after appropriate scaling.

The probability density $p_n$ arises from a determinantal point process,
if there exist (scalar) kernels $K_n$ so that for every $n$ and $k$,
\[ R_{n,k}(x_1, \ldots, x_k) = \det \left[ K_n(x_i,x_j) \right]_{1\leq i, j \leq k}. \]
In particular, one then has for the $1$-point function (particle density),
\[ R_{n,1}(x) = K_n(x,x) \]
and for the probability density $p_n$ itself
\[ p_n(x_1, \ldots, x_n) = \frac{1}{n!} \det \left[ K_n(x_i,x_j) \right]_{1 \leq i,j \leq n}. \]
In this setting the bulk universality comes
down to the statement that the centered and rescaled kernels
\begin{equation} \label{eq:rescaledkernel}
    \frac{1}{c_n} K_n\left(x^* + \frac{x}{c_n}, x^* + \frac{y}{c_n} \right)
    \end{equation}
tend to the sine kernel \eqref{eq:sinekernel} as $n \to \infty$,
and likewise for the edge universality.

Similarly, the probability densities $p_n$ come from Pfaffian point
processes, if there exist  $2\times 2$ matrix kernels $K_n$
so that for every $n$ and $k$,
\[ R_{n,k}(x_1, \ldots, x_k) = \Pf \left[ K_n(x_i,x_j) \right]_{1\leq i, j \leq k}. \]
A proof of universality in a sequence of Pfaffian point processes then comes down
to the proof of a scaling limit for the matrix kernels as $n \to \infty$.

Unitary random matrix ensembles are basic examples of determinantal
point processes, while orthogonal and symplectic  matrix ensembles are examples of
Pfaffian point processes.

For the classical ensembles that are associated with Hermite, Laguerre and Jacobi
polynomials the existence and the form of the limiting kernels has been
proven using the explicit formulas that are available for these classical orthogonal
polynomials, see e.g.\ \cite{For2,Mehta}.

For non-classical ensembles, the results about universality are fairly
complete for unitary ensembles,
due to their connection with orthogonal polynomials. This will be discussed
in more detail in the next section, and we give references there.

The first rigorous results on bulk universality for orthogonal and symplectic
ensembles are due to Stojanovic \cite{Sto} who discussed ensembles with a quartic potential.
This was extended by  Deift and Gioev \cite{DG1} for
 ensembles \eqref{eq:invariantpdf}
with polynomial $V$.
Their techniques are extended to treat edge universality in \cite{DG2,DGKV},
see also the recent monograph \cite{DGbook}. Varying weights are treated
by Shcherbina in \cite{Sh1,Sh2}.

\sect{Unitary random matrix ensembles} \label{sect:unitary}

We explain in more detail how the universality classes arise for the eigenvalues of
a unitary invariant ensemble
\begin{equation} \label{eq:unitaryensemble}
    \frac{1}{\tilde{Z}_{n,N}} e^{- N \Tr V(M)} dM
    \end{equation}
defined on the space of $n \times n$ Hermitian matrices, where
\[ dM = \prod_{i=1}^n d M_{ii}  \prod_{1 \leq i < j \leq n} d \Repart M_{ij} \, d \Impart M_{ij}, \]
and $\tilde{Z}_{n,N}$ is a normalization constant, see also Chapter 4.
The potential function $V$ in \eqref{eq:unitaryensemble}
is typically a polynomial, but could be more general as well. To ensure
that \eqref{eq:unitaryensemble} is well-defined as a probability
measure, we assume that
\begin{equation} \label{eq:growthV}
    \lim_{x \to \pm \infty} \frac{V(x)}{\log(1+x^2)} = +\infty.
    \end{equation}
The factor $N$ in \eqref{eq:unitaryensemble} will typically be
proportional to $n$. This is needed in the large $n$ limit in order
to balance the repulsion among eigenvalues due to the Vandermonde
factor in \eqref{eq:pdfunitary} and the confinement
of eigenvalues due to the potential $V$.

\subsection{Orthogonal polynomial kernel}
The joint probability density for the eigenvalues of a matrix $M$
of the ensemble \eqref{eq:unitaryensemble} has the form
\begin{equation} \label{eq:pdfunitary}
    p_{n,N}(x_1, \ldots, x_n) = \frac{1}{Z_{n,N}}
    \prod_{1 \leq i < j \leq n} (x_i - x_j)^2 \prod_{j=1}^n e^{-N V(x_j)}.
    \end{equation}
Introduce the monic polynomials $P_{k,N}$, $P_{k,N}(x) = x^k + \cdots$,
that are orthogonal with respect to the weight $e^{-NV(x)}$ on $\mathbb R$:
\begin{equation} \label{eq:orthogonality}
    \int_{-\infty}^{\infty}
    P_{k,N}(x) x^j e^{-NV(x)} dx =
        \begin{cases} 0, & \text{for }  j=0, \ldots, k-1, \\
        \gamma_{k,N}^{2} > 0, & \text{for } j = k,
        \end{cases}
        \end{equation}
and the orthogonal polynomial kernel
\begin{equation} \label{eq:OPkernel}
    K_{n,N}(x,y) = \sqrt{e^{-N V(x)}} \sqrt{e^{-N V(y)}}
    \sum_{k=0}^{n-1} \frac{P_{k,N}(x) P_{k,N}(y)}{\gamma_{k,N}^2}.
    \end{equation}
Using the formula for a Vandermonde determinant
and performing elementary row operations we  write \eqref{eq:pdfunitary} as
\begin{align*}
    p_{n,N}(x_1, \ldots, x_n)
    & = \frac{1}{Z_{n,N}} \left(\det \left[ P_{k-1,N}(x_j) \right]_{1 \leq j, k \leq n} \right)^2
        \prod_{j=1}^n e^{-N V(x_j)} \\
    & = \frac{ \prod_{k=0}^{n-1} \gamma_{k,N}^2}{Z_{n,N}}
        \left(\det \left[ \sqrt{e^{-NV(x_j)}} \frac{P_{k-1,N}(x_j)}{\gamma_{k-1,N}} \right]_{1 \leq j,k \leq n}\right)^2.
        \end{align*}
Evaluating the square of the determinant using the rule $\left( \det A \right)^2 = \det (A A^T)$
we obtain that \eqref{eq:pdfunitary} can be written as
\begin{equation} \label{eq:pdfasdet}
    p_{n,N}(x_1, \ldots, x_n) = \frac{1}{n!} \det \left[ K_{n,N}(x_i,x_j) \right]_{1 \leq i, j \leq n},
    \end{equation}
since it can be shown that $Z_{n,N} = n! \, \prod_{k=0}^{n-1} \gamma_{k,N}^2$.
The orthogonality condition \eqref{eq:orthogonality} is then used
to prove that the $k$-point correlation functions \eqref{eq:kpointcorrelation}
are also  determinants
\[ R_{n,k}(x_1, \ldots, x_k) = \det \left[ K_{n,N}(x_i,x_j) \right]_{1 \leq i, j \leq k}, \]
which shows that the eigenvalues are a determinantal point process with correlation kernel $K_{n,N}$.
For the above calculation, see also Chapter 4.

The eigenvalues of orthogonal and symplectic ensembles of random matrices have joint
probability density
\[ \frac{1}{Z_{n,N}} \prod_{1 \leq i < j \leq n} \left| x_i - x_j \right|^{\beta}
    \prod_{j=1}^n e^{-N V(x_j)} \]
with $\beta = 1$ for orthogonal ensembles and $\beta = 4$ for symplectic ensembles.
These probability densities are basic examples of Pfaffian ensembles, as follows
from the calculations in e.g.\ \cite{DGbook,TW2}, see also Chapter 5 of this handbook.

\subsection{Global eigenvalue regime}
As $n, N \to \infty$ such that $n/N \to 1$ the eigenvalues have a limiting
distribution. A weak form of this statement is expressed by the fact that
\begin{equation} \label{eq:macroscopicregime}
    \lim_{\substack{n,N \to \infty \\ n/N \to 1}} \frac{1}{n} K_{n,N}(x,x) = \rho_V(x),
    \qquad x \in \mathbb R
    \end{equation}
exists. The density $\rho_V$ describes the global or macroscopic eigenvalue regime.

The probability measure $d\mu_V(x) = \rho_V(x) dx$  minimizes
the weighted energy
\begin{equation} \label{eq:weightedenergy}
    \iint \log \frac{1}{|x-y|} d\mu(x) d\mu(y) +  \int V(x) d\mu(x)
    \end{equation}
among all Borel probability measures $\mu$ on $\mathbb R$.
Heuristically, it is easy to understand why \eqref{eq:weightedenergy} is relevant.
Indeed, from \eqref{eq:pdfunitary} we find after taking logarithms
and dividing by $-n^2$, that the most likely distribution of eigenvalues
$x_1, \ldots, x_n$ for \eqref{eq:pdfunitary} is the one
that minimizes
\[ \frac{1}{n^2} \sum_{\substack{i,j=1 \\ i \neq j}}^n \log \frac{1}{|x_i-x_j|}
    + \frac{N}{n^2} \sum_{j=1}^n V(x_j). \]
This discrete minimization problem leads to the minimization of
\eqref{eq:weightedenergy} in the continuum limit as $n, N \to \infty$
with $n/N \to 1$.

The minimizer of \eqref{eq:weightedenergy} is unique and
has compact support. It is called
the equilibrium measure in the presence of the external field $V$,
because of its connections with logarithmic potential theory \cite{ST}.
The proof of \eqref{eq:macroscopicregime} with $\rho_V(x) dx$ minimizing
\eqref{eq:weightedenergy} is in \cite{Deift1, Joh}. See also the
remark at the end of subsection \ref{subsec:bulkuniversality} below.
It follows as well from the more general large deviation principle
that is associated with the weighted energy \eqref{eq:weightedenergy}, see \cite{BAG}
and also \cite[\S 2.6]{AGZ}. See also Chapter 14 of this handbook.

If $V$ is real analytic on $\mathbb R$ then $\mu_V$
is supported on a finite union of intervals, say
\[ \supp(\mu_V) = \bigcup_{j=1}^{m} [a_j, b_j], \]
and the density $\rho_V$ takes the form
\begin{equation} \label{eq:rhoV}
    \rho_V(x) = \frac{1}{\pi} h_j(x) \sqrt{(b_j-x)(x-a_j)},
    \qquad x \in [a_j,b_j], \quad j = 1, \ldots, m
    \end{equation}
where $h_j$ is real analytic and non-negative on $[a_j,b_j]$, see \cite{DKM}.
Another useful representation is that
\begin{equation} \label{eq:rhoV2}
    \rho_V(x) = \frac{1}{\pi} \sqrt{q_V^-(x)},
    \qquad x \in \mathbb R,
    \end{equation}
where $q_V^-$ is the negative part of
\begin{equation} \label{eq:qV}
    q_V(x) = \left( \frac{V'(x)}{2} \right)^2 - \int \frac{V'(x)-V'(s)}{x-s} d\mu_V(s).
    \end{equation}
If $V$ is a polynomial then $q_V$ is a polynomial of degree $2(\deg V - 1)$.
In that case the number of intervals in the support is bounded by $\frac{1}{2} \deg V$,
see again \cite{DKM}. See also a discussion in Chapters 14 and 16.

\subsection{Local eigenvalue regime}
In the context of Hermitian matrix models, the universality may be stated
as the fact that the global eigenvalue regime
determines the local eigenvalue regime.
The universality results take on a different form depending on the
nature of the reference point $x^*$.

\paragraph{Bulk universality}
A regular point in the bulk is an interior point $x^*$ of $\supp(\mu_V)$
 such that $\rho_V(x^*) > 0$. At a regular point in the bulk one has
\begin{equation} \label{eq:bulkuniversality}
    \lim_{n \to \infty} \frac{1}{cn} K_{n,N} \left( x^* + \frac{x}{cn}, x^* + \frac{y}{cn} \right)
    = K^{\sin}(x,y) \end{equation}
    where $c = \rho_V(x^*)$ and $K^{\sin}$ is the sine kernel \eqref{eq:sinekernel}.

Convincing heuristic arguments for the universality of the sine kernel
were given by Br\'ezin and Zee \cite{BZ} by the method
of orthogonal polynomials. Supersymmetry arguments were used
in \cite{HW}.
Rigorous results for bulk universality \eqref{eq:bulkuniversality} beyond
the classical ensembles were first given by Pastur and Shcherbina \cite{PS1},
and later by Bleher and Its \cite{BI1} and by Deift et al.\ \cite{DKMVZ1,DKMVZ2}
for real analytic $V$.
In these papers the Riemann-Hilbert techniques were introduced in the study
of orthogonal polynomials, which we will review in section \ref{sect:RHmethod} below
Conditions on $V$ in \eqref{eq:invariantpdf2}
were further weakened in \cite{MM,PS2}.

In more recent work of Lubinsky \cite{Lub1,Lub3,Lub4} and Levin and Lubinsky \cite{LL1,LL2}
it was shown that bulk universality holds under extremely weak
conditions. One of the results is that in an ensemble \eqref{eq:invariantpdf}
restricted to a compact interval, bulk universality holds at each
interior point where $V$ is continuous.

\paragraph{Soft edge universality}
An edge point $x^* \in \{a_1, b_1, \ldots, a_m, b_m\}$ is a regular
edge point of $\supp(\mu_V)$ if $h_j(x^*) > 0$ in \eqref{eq:rhoV}.
In that case the density $\rho_V$ vanishes as a square root at $x^*$.
The scaling limit is then the Airy kernel \eqref{eq:Airykernel}.
If $x^* = b_j$ is a right-edge point then for a certain constant $c > 0$,
\begin{equation} \label{eq:softedgeuniversality}
    \lim_{n \to \infty} \frac{1}{(cn)^{2/3}} K_{n,N} \left( x^* + \frac{x}{(cn)^{2/3}},
    x^* + \frac{y}{(cn)^{2/3}} \right)
        = K^{\Ai}(x,y)
        \end{equation}
while if $x^* = a_j$ is a left-edge point, we find the same limit after a change
of sign
\begin{equation}
    \lim_{n \to \infty} \frac{1}{(cn)^{2/3}} K_{n,N} \left( x^* - \frac{x}{(cn)^{2/3}},
    x^* - \frac{y}{(cn)^{2/3}} \right) \\
        = K^{\Ai}(x,y).
        \end{equation}
Thus the regular edge points belong to the Airy universality class.

The Airy kernel was implicitly derived in \cite{BB}. For classical ensembles
the soft edge universality \eqref{eq:softedgeuniversality}
is derived in \cite{For1,NW}, see also \cite{For2}.
For quartic and sextic potentials $V$, it is derived in \cite{KF1} by
the so-called Shohat method.
For real analytic potentials potentials it is implicit in \cite{DKMVZ2}
and made explicit in \cite{DG2}.

\paragraph{Hard edge universality}
The Bessel universality classes arise for eigenvalues of positive definite matrices.
Let $\alpha > -1$ be a parameter and consider
\[ \frac{1}{\widetilde{Z}_{n,N}} (\det M)^{\alpha} e^{-N \Tr V(M)} dM \]
as a probability measure on the space of $n \times n$ Hermitian positive definite matrices. For
the case $V(x) = x$
this is the Laguerre Unitary Ensemble, which is also known as a complex Wishart ensemble,
see Chapter 4.

Assuming that $\alpha$ remains fixed, the global eigenvalue regime
does not depend on $\alpha$.
The eigenvalue density $\rho_V$ lives on $[0,\infty)$, and for $x \in [0,\infty)$ it
continues to have the representation \eqref{eq:rhoV2}
but now $q_V$ is modified to
\begin{equation} \label{eq:qVhard}
    q_V(x) = \left( \frac{V'(x)}{2} \right)^2 - \int \frac{V'(x)-V'(s)}{x-s} d\mu_V(s)
    - \frac{1}{x} \int V'(s) d\mu_V(s).
    \end{equation}
If $\int V'(s) d\mu_V(s) > 0$ then $0$ is in the support
of $\mu_V$ and $\rho_V$ has a square-root singularity at $x=0$.

The effect of the parameter $\alpha$ is noticeable in the local eigenvalue regime
near $x^*=0$. If $\int V'(s) d\mu_V(s) > 0$, then the limiting kernel is the
Bessel kernel \eqref{eq:Besselkernel} of order $\alpha$. For $x, y > 0$ and
for an appropriate constant $c > 0$, we have
\begin{equation} \label{eq:hardedgeuniversality}
    \lim_{n \to \infty} \frac{1}{(cn)^2} K_{n,N} \left( \frac{x}{(cn)^2}, \frac{y}{(cn)^2} \right)
    = K^{\Bes,\alpha}(x,y). \end{equation}
The hard edge universality \eqref{eq:hardedgeuniversality} was
proven in \cite{KV1,Lub2}.

\paragraph{Spectral singularity}

A related class of Bessel kernels
\begin{equation} \label{eq:originkernels}
    \widehat{K}^{\Bes,\alpha}(x,y) = \pi \sqrt{x} \sqrt{y}
    \frac{J_{\alpha+\frac{1}{2}}(\pi x) J_{\alpha-\frac{1}{2}}(\pi y) -
    J_{\alpha-\frac{1}{2}}(\pi x) J_{\alpha+\frac{1}{2}}(\pi y)}{2(x-y)}
    \end{equation}
 appears as scaling limits in Hermitian matrix models of the form
\begin{equation} \label{eq:spectralsingularitymodel}
    \frac{1}{\tilde{Z}_{n,N}} \left| \det M \right|^{2\alpha}
     e^{-N \Tr V(M)} dM \end{equation}
with $\alpha > -1/2$.

The extra factor $\left| \det M \right|^{2\alpha}$ is referred to
as a spectral singularity and the matrix model \eqref{eq:spectralsingularitymodel}
is relevant in quantum chromodynamics, where it is referred to as a chiral ensemble,
see Chapter 32 of this handbook.
The spectral singularity does not
change the global density $\rho_V$ of eigenvalues, but it does
have an influence on the local eigenvalue statistics at the origin.
Assuming that $c = \rho_V(0) > 0$, one now finds
\begin{equation} \label{eq:spectralsingularity}
    \lim_{n \to \infty} \frac{1}{cn} K_{n,N} \left(\frac{x}{cn}, \frac{y}{cn} \right)
    = \widehat{K}^{\Bes,\alpha}(x,y), \qquad x, y > 0, \end{equation}
instead of \eqref{eq:bulkuniversality}.

The universality of the Bessel kernels \eqref{eq:Besselkernel} and \eqref{eq:originkernels} was
discussed by Akemann et al.\ \cite{ADMN1} for integer values of $\alpha$. It
was extended to non-integer $\alpha$ in \cite{KF2}. The model \eqref{eq:spectralsingularitymodel}
was analyzed in \cite{KV2} with the Riemann-Hilbert method.

See \cite{KlV} for the spectral universality in orthogonal
ensembles.

\paragraph{Weight with jump discontinuity}

More recently \cite{FMS}, a new class of limiting kernels was
identified for unitary ensembles of the form
\[ \frac{1}{\tilde{Z}_{n}} e^{- \Tr V(M)}  dM \]
defined on Hermitian matrices with eigenvalues in $[-1,1]$
in cases where the weight function $w(x) = e^{-V(x)}$, $x \in [-1,1]$,
has a jump discontinuity at $x=0$.
The limiting kernels in \cite{FMS} are constructed
out of confluent hypergeometric functions, see also \cite{IK}.

\sect{Riemann-Hilbert method} \label{sect:RHmethod}

A variety of methods have been developed to prove the above universality results
in varying degrees of generality. One of these methods will be described here,
namely the steepest descent analysis for the Riemann-Hilbert problem (RH problem).
This is a powerful method to obtain strong and uniform asymptotics for orthogonal
polynomials. One of the outcomes is the limiting behavior of the
eigenvalue correlation kernels. However, the method gives much more.
It is also able to give asymptotics of the recurrence
coefficients, Hankel determinants and other notions associated
with orthogonal polynomials. In this section we closely follow the
paper \cite{DKMVZ2} of Deift et al, see also \cite{Deift1}.

\subsection{Statement of the RH problem}
A Riemann-Hilbert problem is a jump problem for a piecewise analytic
function in the complex plane. The RH problem for orthogonal
polynomials asks for a $2 \times 2$ matrix valued function $Y$ satisfying
\begin{itemize}
\item[(1)] $Y : \mathbb C \setminus \mathbb R \to \mathbb C^{2\times 2}$ is analytic,
\item[(2)] $Y$ has boundary values on the real line, denoted by $Y_{\pm}(x)$,
where $Y_+(x)$ ($Y_-(x)$) denotes the limit of $Y(z)$ as $z \to x \in \mathbb R$
with $\Impart z > 0$ ($\Impart z < 0$), and
\[ Y_+(x) = Y_-(x) \begin{pmatrix} 1 & e^{-NV(x)} \\ 0 & 1 \end{pmatrix}, \]
\item[(3)] $Y(z) = (I + O(1/z)) \begin{pmatrix} z^n & 0 \\ 0 & z^{-n} \end{pmatrix}$
    as $z \to \infty$.
    \end{itemize}
The unique solution is given in terms of the orthogonal polynomials
\eqref{eq:orthogonality} by
\begin{equation} \label{eq:solutionY}
    Y(z) =
    \begin{pmatrix} \ds P_{n,N}(z) &
    \ds \frac{1}{2\pi i} \int_{-\infty}^{\infty} \frac{P_{n,N}(x) e^{-NV(x)}}{x-z} \, dx \\[10pt]
    \ds -2\pi i \gamma_{n-1,N}^2 P_{n-1,N}(z) &
    \ds - \gamma_{n-1,N}^2 \int_{-\infty}^{\infty} \frac{P_{n-1,N}(x) e^{-NV(x)}}{x-z} \, dx
    \end{pmatrix}.
    \end{equation}
The RH problem for orthogonal polynomials and its solution \eqref{eq:solutionY}
are due to Fokas, Its, and Kitaev \cite{FIK}.

By the Christoffel-Darboux formula for orthogonal polynomials,
we have that the orthogonal polynomial kernel
\eqref{eq:OPkernel} is equal to
\begin{multline} \label{eq:OPkernelCD}
    K_{n,N}(x,y) = \\ \sqrt{e^{-N V(x)}} \sqrt{e^{-N V(y)}} \, \gamma_{n-1,N}^2 \,
        \frac{P_{n,N}(x) P_{n-1,N}(y) - P_{n-1,N}(x) P_{n,N}(y)}{x-y} \end{multline}
which in view of \eqref{eq:solutionY} and the fact that $\det Y(z) \equiv 1$ can be rewritten as
\begin{multline} \label{eq:kernelinY}
    K_{n,N}(x,y) = \\
     \frac{1}{2\pi i (x-y)} \sqrt{e^{-NV(x)}} \sqrt{e^{-NV(y)}}
        \begin{pmatrix} 0 & 1 \end{pmatrix} Y_+(y)^{-1} Y_+(x) \begin{pmatrix} 1 \\ 0 \end{pmatrix},
        \end{multline}
for $x, y \in \mathbb R$.
Other notions related to the orthogonal polynomials are also contained in the
solution of the Riemann-Hilbert problem.
The monic polynomials satisfy a three-term recurrence relation
\[ x P_{n,N}(x) = P_{n+1,N}(x) + b_{n,N} P_{n,N}(x) + a_{n,N} P_{n-1,N}(x) \]
with recurrence coefficients $a_{n,N} > 0$ and $b_{n,N} \in \mathbb R$.
The recurrence coefficients can be found from the solution of the RH problem
by expanding $Y$ around $\infty$:
\[ Y(z) = \left(I + \frac{1}{z} Y_1 + \frac{1}{z^2} Y_2 + \cdots \right)
    \begin{pmatrix} z^{n} & 0 \\ 0 & z^{-n} \end{pmatrix} \]
as $z \to \infty$. The $2 \times 2$ matrices $Y_1$ and $Y_2$ do not depend on $z$,
but do depend on $n$ and $N$. Then
\begin{align}
    a_{n,N}  = \left(Y_1\right)_{12} \left(Y_1\right)_{21}, \qquad
    b_{n,N}  = \frac{(Y_2)_{12}}{(Y_1)_{12}} - \left(Y_1\right)_{22}.
    \end{align}

\subsection{Outline of the steepest descent analysis}
The steepest descent analysis of RH problems is due to Deift
and Zhou \cite{DZ}.
It produces a number of explicit and invertible transformations
leading to a RH problem for a new matrix-valued function $R$
with identitiy asymptotics at infinity. Also $R$ depends on $n$ and $N$,
and as $n,N \to \infty$ with $n/N \to 1$, the jump matrices tend
to the identity matrix,  in regular cases typically at a rate of $O(1/n)$.
Then it can be shown that $R(z)$ tends to the identity matrix
as $n, N \to \infty$ with $n/N \to 1$ and also at a rate of $O(1/n)$
in regular cases.

Following the transformations in the steepest descent analysis we can
then find asymptotic formulas for $Y$ and in particular for the orthogonal
polynomial $P_{n,N}$, the recurrence coefficients and the correlation kernel.

With more work one may be able to obtain more precise asymptotic information
on $R$. For example, if $n = N$ and if we are in the one-cut regular case, then
there is an asymptotic expansion for $R(z)$:
\[ R(z) \sim I + \frac{R^{(1)}(z)}{n} + \frac{R^{(2)}(z)}{n^2} + \cdots \]
with explicitly computable matrices $R^{(j)}(z)$. This in turn leads to
asymptotic expansions for the orthogonal polynomials as well.
We will not go into this aspect here.

Here we present the typical steps in the Deift-Zhou steepest descent analysis.
We focus on the one-cut case, that is on the situation where $V$
is real analytic and the equilibrium measure $\mu_V$ is supported on one interval $[a,b]$.
We also assume that we are in a regular case, which means that
\[ \rho_V(x) = \frac{d\mu_V(x)}{dx} = \frac{1}{\pi} h(x) \sqrt{(b-x)(x-a)}, \qquad x \in [a,b], \]
with $h(x) > 0$ for $x \in [a,b]$, and strict inequality holds in
the variational condition \eqref{eq:varcondition2} below.
Generically these regularity conditions are satisfied, see \cite{KMc}. The singular cases lead
to different universality classes, and this will be commented on in
section \ref{sect:non-standard}.

For convenience we also take $N = n$.

\subsection{First transformation: normalization  at infinity}

The equilibrium measure $\mu_V$ is used in the first transformation
of the RH problem.
The equilibrium measure satisfies for some constant $\ell$
\begin{align} \label{eq:varcondition1}
    2 \int \log \frac{1}{|x-y|} d\mu_V(y) + V(x) & = \ell, \qquad x \in [a,b], \\
    2 \int \log \frac{1}{|x-y|} d\mu_V(y) + V(x) & \geq \ell, \qquad x \in \mathbb R \setminus [a,b].
    \label{eq:varcondition2}
    \end{align}
These are the Euler-Lagrange variational conditions associated with
the minimization of the weighted energy \eqref{eq:weightedenergy}.
Since \eqref{eq:weightedenergy} is a strictly convex functional on
finite energy probability measures, the conditions \eqref{eq:varcondition1}--\eqref{eq:varcondition2}
characterize the minimizer $\mu_V$.

The equilibrium measure $\mu_V$ leads to the $g$-function
\begin{equation} \label{eq:defg}
    g(z) = \int \log(z-x) \, d\mu_V(x), \qquad z \in \mathbb C \setminus (-\infty,b],
    \end{equation}
which is used in the first transformation $Y \mapsto T$.
We put
\begin{equation} \label{eq:defT}
    T(z) = \begin{pmatrix} e^{-n \ell/2} & 0 \\ 0 & e^{n \ell/2} \end{pmatrix}
    Y(z) \begin{pmatrix} e^{-n (g(z) - \ell/2)} & 0 \\ 0 & e^{n(g(z) - \ell/2)} \end{pmatrix}.
    \end{equation}

The jumps in the RH problem for $T$ are conveniently stated in terms of
the functions
\begin{align} \label{eq:defphi}
    \phi(z) & =  \int_b^z h(s) ((s-b)(s-a))^{1/2} \, ds \\
    \widetilde{\phi}(z) & = \int_a^z h(s) ((s-b)(s-a))^{1/2} \, ds,
    \label{eq:defvarphi}
    \end{align}
where we use the fact that $V$ is real analytic and
therefore $h$ has an  analytic continuation to a neighborhood
of the real line.
Then $T$ satisfies the RH problem
\begin{itemize}
\item[(1)] $T$ is analytic in $\mathbb C \setminus \mathbb R$.
\item[(2)] On $\mathbb R$ we have the jump $T_+ = T_- J_T$
where
\begin{equation} \label{eq:defJT}
    J_T(x) = \begin{cases} \begin{pmatrix} e^{2n \phi_+(x)} & 1 \\ 0 & e^{2n\phi_-(x)} \end{pmatrix}
    & \quad x \in (a,b), \\
    \begin{pmatrix} 1 & e^{-2n \phi(x)} \\ 0 & 1 \end{pmatrix}
    & \quad x > b, \\
    \begin{pmatrix} 1 & e^{-2n \widetilde{\phi}(x)} \\ 0 & 1 \end{pmatrix}
    & \quad x < a.
    \end{cases} \end{equation}
\item[(3)] $T(z) = I + O(1/z)$ as $ z \to \infty$.
\end{itemize}

The RH problem is now normalized at infinity. The jump matrices on $(-\infty, a)$
and $(b,\infty)$ tend to the identity matrix as $n \to \infty$, since $\phi(x) > 0$
for $x > b$ and $\widetilde{\phi}(x) > 0$ for $x < a$, which is due to the assumption
that strict inequality holds in \eqref{eq:varcondition2}.
We have to deal with the jump on $(a,b)$. The diagonal entries
in the jump matrix on $(a,b)$ are highly oscillatory. The goal of the next transformation is
to turn these highly oscillatory entries into exponentially decaying ones.

\subsection{Second transformation: opening of lenses}

\begin{figure}[t]
\centering
\unitlength 1pt
\linethickness{0.5pt}
\begin{picture}(240,100)(-100,-50)
   \put(0,0){\line(1,0){140}}
   \put(0,0){\line(-1,0){120}}
   \qbezier(-50,0)(-5,50)(40,0)
   \qbezier(-50,0)(-5,-50)(40,0)
   \qbezier(80,0)(79,1)(75,5)   \qbezier(80,0)(79,-1)(75,-5)
   \qbezier(-10,0)(-11,1)(-15,5)   \qbezier(-10,0)(-11,-1)(-15,-5)
   \qbezier(-5,25)(-6,26)(-10,30)   \qbezier(-5,25)(-6,24)(-10,20)
   \qbezier(-5,-25)(-6,-26)(-10,-30)   \qbezier(-5,-25)(-6,-24)(-10,-20)
   \qbezier(-80,0)(-81,1)(-85,5)   \qbezier(-80,0)(-81,-1)(-85,-5)
   \put(-52,-10){$a$}
   \put(-50,0){\circle*{4}}
   \put(42,-10){$b$}
   \put(40,0){\circle*{4}}
   \put(80,15){$\begin{pmatrix} 1 & e^{-2n\phi}  \\ 0 & 1\end{pmatrix}$}
   \put(40,-40){$\begin{pmatrix} 0 & 1 \\  -1 & 0 \end{pmatrix}$}
   \put(0,35){$\begin{pmatrix} 1 & 0 \\ e^{2n\phi} & 1 \end{pmatrix}$}
   \put(-70,-40){$\begin{pmatrix} 1 & 0 \\ e^{2n\phi} & 1 \end{pmatrix}$}
   \put(-120,15){$\begin{pmatrix} 1 & e^{-2n\widetilde{\phi}} \\ 0 & 1 \end{pmatrix}$}
   \put(45,-25){\vector(-2,1){50}}
   \end{picture}
   \caption{Contour $\Sigma_S$ and jump matrices for the  RH problem for $S$.
   This figure is reproduced from  \cite{KT}.}\label{figure1}
\end{figure}
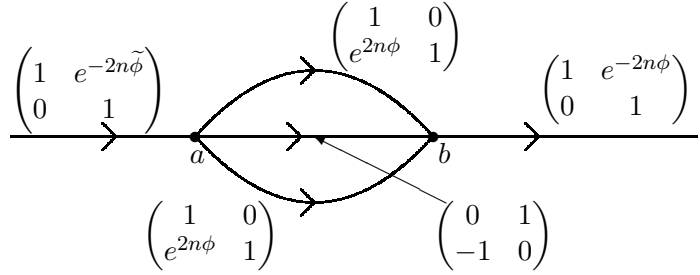

We open up a lens around the interval $[a,b]$ as in Figure \ref{figure1}
and define
\begin{align} \label{eq:defS}
    S = \begin{cases}
        T \begin{pmatrix} 1 & 0 \\ - e^{2n\phi} & 0 \end{pmatrix}
        & \textrm{in the upper part of the lens}, \\
        T \begin{pmatrix} 1 & 0 \\  e^{2n\phi} & 0 \end{pmatrix}
        & \textrm{in the lower part of the lens} \\
        T & \textrm{elsewhere}.
        \end{cases}
        \end{align}
Then $S$ is defined and analytic in $\mathbb C \setminus \Sigma_S$
where $\Sigma_S$ is the contour consisting of the real line, and the
upper and lower lips of the lens, with orientation as indicated by
the arrows in Figure \ref{figure1}. The orientation induces
a $+$-side and a $-$-side on each part of $\Sigma_S$, where the $+$-side is on
the left as one traverses the contour according to its orientation
and the $-$-side is on the right.
We use $S_{\pm}$ to denote the limiting values of $S$ on $\Sigma_S$ when
approaching $\Sigma_S$ from the $\pm$-side. This convention about $\pm$-limits,
depending on the orientation of the contour is usual in Riemann-Hilbert
problems and will also be used later on.

Then $S$ satisfies the following RH problem:
\begin{itemize}
\item[(1)] $S$ is analytic in $\mathbb C \setminus (\mathbb R \cup \Sigma_S)$.
\item[(2)] On $\Sigma_S$ we have the jump $S_+ = S_- J_S$
where
\begin{equation} \label{eq:defJS}
    J_S(x) = \begin{cases}
    \begin{pmatrix} 0 & 1 \\ -1 & 0 \end{pmatrix}
    & \textrm{for } x \in (a,b), \\
    \begin{pmatrix} 1 & 0 \\ e^{2n\phi(x)} & 1 \end{pmatrix}
    & \textrm{on the lips of the lens}, \\
    J_T(x) & \textrm{for } x < a \textrm{ or } x > b.
    \end{cases}
    \end{equation}
\item[(3)] $S(z) = I + O(1/z)$ as $z \to \infty$.
\end{itemize}

If the lens is taken sufficiently small, we can guarantee that
$\Repart \phi < 0$ on the lips of the lens. Then the jump matrices
for $S$ tend to the identity matrix as $n \to \infty$ on the lips
of the lens and on the unbounded intervals $(-\infty, a)$ and $(b, \infty)$.

\subsection{Outside parametrix}

The next step is to build an approximation to $S$, valid for large $n$,
the so-called parametrix. The parametrix consists of two parts, an
outside or global parametrix that will model $S$ away from the endpoints
and local parametrices that are good approximation to $S$ in a neighborhood
of the endpoints.

The outside parametrix $M$ satisfies
\begin{itemize}
\item[(1)] $M$ is analytic in $\mathbb C \setminus [a,b]$,
\item[(2)] $M$ has the jump
\[ M_+(x) = M_-(x) \begin{pmatrix} 0 & 1 \\ - 1 & 0 \end{pmatrix}
    \quad \text{for } x \in (a,b). \]
\item[(3)] $M(z) = I + O(1/z)$ as $z \to \infty$.
\end{itemize}
To have uniqueness of a solution we also impose
\begin{itemize}
\item[(4)] $M$ has at most fourth-root singularities at the
endpoints $a$ and $b$.
\end{itemize}

The solution to this RH problem is given by
\begin{equation} \label{eq:defM}
    M(z) = \begin{pmatrix} \ds \frac{\beta(z) + \beta^{-1}(z)}{2} & \ds \frac{\beta(z) - \beta^{-1}(z)}{2i} \\[10pt]
    - \ds \frac{\beta(z) - \beta^{-1}(z)}{2i} & \ds \frac{\beta(z) + \beta^{-1}(z)}{2}
    \end{pmatrix},
        \qquad z \in \mathbb C \setminus [a,b],
        \end{equation}
where
\begin{equation} \label{eq:defbeta}
    \beta(z) = \left(\frac{z-b}{z-a} \right)^{1/4}.
    \end{equation}

\subsection{Local parametrix}
The local parametrix is constructed in neighborhoods
$U_{\delta}(a) = \{ z \mid |z-a| < \delta\}$ and
$U_{\delta}(b) = \{ z \mid |z-b| < \delta\}$ of the endpoints
$a$ and $b$, where $\delta > 0$ is small, but fixed.

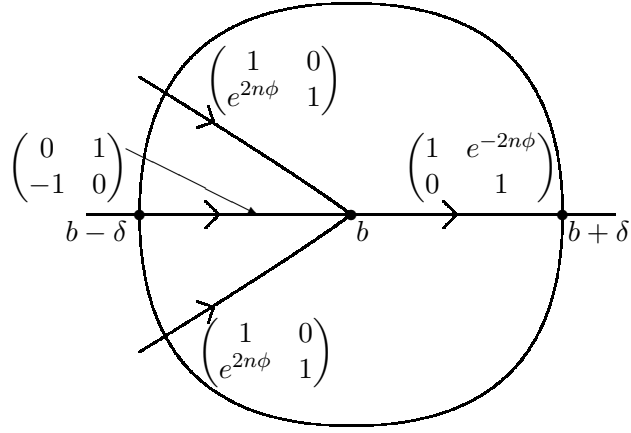
\begin{figure}[t]
\centering
\unitlength 1pt
\linethickness{0.5pt}
\begin{picture}(240,200)(-100,-100)
   \put(-60,0){\line(1,0){200}}
   \qbezier(-40,52)(20,15)(40,0)
   \qbezier(-40,-52)(20,-15)(40,0)
   \qbezier(80,0)(79,1)(75,5)   \qbezier(80,0)(79,-1)(75,-5)
   \qbezier(-10,0)(-11,1)(-15,5)   \qbezier(-10,0)(-11,-1)(-15,-5)
   \qbezier(-12,35)(-13,38)(-14,41)   \qbezier(-12,35)(-15,34)(-18,33)
   \qbezier(-12,-35)(-13,-38)(-14,-41)   \qbezier(-12,-35)(-15,-34)(-18,-33)
   \put(42,-10){$b$}
   \put(40,0){\circle*{4}}
   \put(120,0){\circle*{4}}
   \put(-40,0){\circle*{4}}
   \put(122,-10){$b+\delta$}
   \put(-68,-10){$b-\delta$}
   \qbezier(120,0)(120,80)(40,80)
   \qbezier(40,80)(-40,80)(-40,0)
   \qbezier(40,-80)(-40,-80)(-40,0)
   \qbezier(120,0)(120,-80)(40,-80)
   \put(60,15){$\begin{pmatrix} 1 & e^{-2n\phi}  \\ 0 & 1\end{pmatrix}$}
   \put(-90,15){$\begin{pmatrix} 0 & 1 \\  -1 & 0 \end{pmatrix}$}
   \put(-15,48){$\begin{pmatrix} 1 & 0 \\ e^{2n\phi} & 1 \end{pmatrix}$}
   \put(-18,-55){$\begin{pmatrix} 1 & 0 \\ e^{2n\phi} & 1 \end{pmatrix}$}
   \put(-45,25){\vector(2,-1){50}}
   \end{picture}
   \caption{Neighborhood $U_{\delta}(b)$ of $b$ and contours and jump matrices for the RH problem for $P$}
   \label{figure2}.
\end{figure}

In $U_{\delta}(a) \cup U_{\delta}(b)$
we want to have a $2 \times 2$ matrix valued function $P$ satisfying
the following
(see also Figure \ref{figure2} for the jumps
in the neighborhood of $b$):
\begin{itemize}
\item[(1)] $P$ is defined and analytic in $\left( U_{\delta}(a) \cup U_{\delta}(b) \right) \setminus \Sigma_S$
and has a continuous extension to $\left( \overline{U_{\delta}(a)} \cup \overline{U_{\delta}(b)} \right)
\setminus \Sigma_S$.
\item[(2)] On $\Sigma_S \cap \left( U_{\delta}(a) \cup U_{\delta}(b) \right)$ there is the jump
\[ P_+ = P_- J_S \]
where $J_S$ is the jump matrix \eqref{eq:defJS} in the RH problem for $S$.
\item[(3)] $P$ agrees with the global parametrix $M$ on the boundaries of
$U_{\delta}(a)$ and $U_{\delta}(b)$ in the sense that
\begin{equation} \label{eq:matching}
    P(z) = \left(I + O\left(\frac{1}{n}\right)\right) M(z)
    \end{equation}
    as $n \to \infty$ uniformly for $z \in \partial U_{\delta}(a) \cup \partial U_{\delta}(b)$.
\item[(4)] $P(z)$ remains bounded as $z \to a$ or $z \to b$.
\end{itemize}

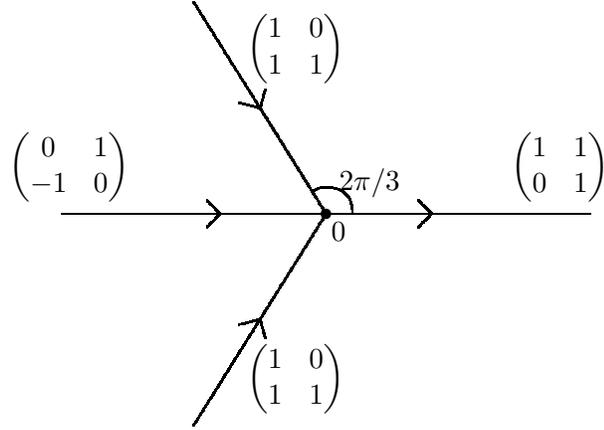
\begin{figure}[t]
\centering
\unitlength 1pt
\linethickness{0.5pt}
\begin{picture}(200,160)(-100,-80)
   \put(0,0){\line(1,0){100}}
   \put(0,0){\line(-1,0){100}}
   \qbezier(0,0)(-25,40)(-50,80)
   \qbezier(0,0)(-25,-40)(-50,-80)
   \qbezier(40,0)(39,1)(35,5)   \qbezier(40,0)(39,-1)(35,-5)
   \qbezier(-40,0)(-41,1)(-45,5)   \qbezier(-40,0)(-41,-1)(-45,-5)
   \qbezier(-25,40)(-24,43.5)(-23,47)   \qbezier(-25,40)(-29,41)(-33,42)
   \qbezier(-25,-40)(-24,-43.5)(-23,-47)   \qbezier(-25,-40)(-29,-41)(-33,-42)
   \put(2,-10){$0$}
   \put(0,0){\circle*{4}}
   \qbezier(10,0)(10,10)(0,10)
   \qbezier(0,10)(-4,10)(-5,8)
   \put(5,9){$2\pi/3$}
   \put(70,15){$\begin{pmatrix} 1 & 1  \\ 0 & 1\end{pmatrix}$}
   \put(-30,60){$\begin{pmatrix} 1 & 0 \\  1 & 1 \end{pmatrix}$}
   \put(-120,15){$\begin{pmatrix} 0& 1  \\ -1 & 0\end{pmatrix}$}
   \put(-30,-65){$\begin{pmatrix} 1 & 0 \\ 1 & 1 \end{pmatrix}$}
   \end{picture}
   \caption{Contour $\Sigma_A$ and jump matrices for the Airy Riemann-Hilbert problem\label{figure3}.}
\end{figure}

The solution of the RH problem for $P$ is constructed out of
the Airy function
\[ y_0(z) = \Ai(z), \]
and its rotated versions
\[ y_1(z) = \omega \Ai(\omega z), \qquad y_2(z) = \omega^2 \Ai(\omega^2 z),
    \qquad \omega = e^{2\pi i/3}. \]
These three solutions of the Airy differential equation
$ y''(z) = z y(z)$
are connected by the identity
\[ y_0(z) + y_1(z) + y_2(z) = 0. \]

They are used in the solution of the following model RH problem
posed on the contour $\Sigma_A$ shown in Figure \ref{figure3}.
\begin{itemize}
\item[(1)] $A : \mathbb C \setminus \Sigma_A \to \mathbb C^{2\times 2}$ is analytic.
\item[(2)] $A$ has jump $A_+ = A_- J_A$
on $\Sigma_A$ with jump matrix $J_A$ given by
\begin{align}
    J_A(z)  = \begin{cases}
        \begin{pmatrix} 1 & 1 \\ 0 & 1 \end{pmatrix}
        & \textrm{for } \arg z = 0, \\
        \begin{pmatrix} 1 & 0 \\ 1 & 1 \end{pmatrix}
        & \textrm{for } \arg z = \pm 2 \pi i/3,  \\
        \begin{pmatrix} 0 & 1 \\ -1 & 0 \end{pmatrix}
        & \textrm{for } \arg z  = \pi.
        \end{cases}
    \end{align}
\item[(3)] As $z \to \infty$, we have
\begin{equation} \label{eq:Aasymptotics}
    A(z) =
    \begin{pmatrix} z^{-1/4} & 0 \\ 0 & z^{1/4} \end{pmatrix}
    \frac{1}{\sqrt{2}} \begin{pmatrix} 1 & i \\ i & 1 \end{pmatrix}
    \left(I + O(z^{-3/2})\right)
    \begin{pmatrix}
    e^{- \frac{2}{3} z^{3/2}} & 0 \\
    0 & e^{\frac{2}{3} z^{3/2}} \end{pmatrix}.
    \end{equation}
\item[(4)] $A(z)$ remains bounded as $z \to 0$.
\end{itemize}

The rather complicated asymptotics in \eqref{eq:Aasymptotics} corresponds
to the asymptotic formulas
\begin{equation} \label{eq:Airyasymptotics}
\begin{aligned}
    \Ai(z) & = \frac{1}{2\sqrt{\pi} z^{1/4}} e^{-\frac{2}{3} z^{3/2}}
    \left( 1 + \mathcal O(z^{-3/2})\right), \\
    \Ai'(z) & = - \frac{z^{1/4}}{2\sqrt{\pi}} e^{-\frac{2}{3} z^{3/2}}
    \left( 1 + \mathcal O(z^{-3/2}) \right),
    \end{aligned}
\end{equation}
as $z \to 0$ with $-\pi < \arg z < \pi$, that are known for the Airy function and its derivative.

The solution of the Airy Riemann-Hilbert problem is as follows
\begin{equation} \label{eq:defA}
    A(z) = \sqrt{2 \pi} \times
        \begin{cases} \begin{pmatrix} y_0(z) & - y_2(z) \\
    -i y_0'(z)  & i y_2'(z) \end{pmatrix},
     & 0 < \arg z < 2\pi/3, \\
        \begin{pmatrix} - y_1(z) & -y_2(z) \\
       -i y_1'(z)  & i y_2'(z) \end{pmatrix},
       & 2\pi/3 < \arg z < \pi, \\
         \begin{pmatrix} - y_2(z) & y_1(z) \\
    i y_2'(z) & -i y_1'(z)  \end{pmatrix},
    & -\pi  < \arg z < -2\pi/3, \\
     \begin{pmatrix} y_0(z) & y_1(z)  \\
    -i y_0'(z)  & -i y_1'(z)  \end{pmatrix},
    &  -2 \pi/3 < \arg z < 0.
        \end{cases}
        \end{equation}
The constants $\sqrt{2\pi}$ and $\pm i$ are such that
$\det A (z) \equiv 1$ for $z \in \mathbb C \setminus \Sigma_A$.

To construct the local parametrix $P$ in the neighborhood
$U_{\delta}(b)$ of $b$ we also need the function
\[ f(z) = \left[ \frac{3}{2} \phi(z) \right]^{2/3} \]
which is a conformal map from $U_{\delta}(b)$ (provided $\delta$
is small enough) onto a neighborhood of the origin. For this it is important
that the density $\rho_V$ vanishes as a square root at $b$.
We may assume that the lens around $(a,b)$ is opened in such a way that
the parts of the lips of the lens within
$U_{\delta}(b)$ are mapped by $f$ into the rays $\arg
z = \pm 2 \pi/3$.

Then the local parametrix $P$ is given in $U_{\delta}(b)$ by
\begin{equation} \label{eq:defP}
    P(z) = E_n(z) A(n^{2/3} f(z)) \begin{pmatrix} e^{n \phi(z)} & 0 \\ 0 & e^{-n \phi(z)} \end{pmatrix},
    \qquad z \in U_{\delta}(b) \setminus \Sigma_S,
    \end{equation}
where the  prefactor $E_n(z)$ is given explicitly by
\begin{multline} \label{eq:prefactorE}
     E_n(z)
        = -\frac{1}{\sqrt{2}} \begin{pmatrix} 1 & -i\\ -i & 1 \end{pmatrix}
        \begin{pmatrix} n^{1/6} (z-a)^{1/4} & 0 \\ 0 & n^{-1/6} (z-a)^{-1/4} \end{pmatrix} \\
       \times \begin{pmatrix}  \left(f(z)/(z-b) \right)^{1/4} & 0 \\
        0 &  \left(f(z)/(z-b) \right)^{-1/4}
        \end{pmatrix}.
    \end{multline}
Then $E_n$ is analytic in $U_{\delta}(b)$ and it does not change the jump conditions.
It is needed in order to satisfy the matching condition \eqref{eq:matching}
on the boundary $|z-b| = \delta$ of $U_{\delta}(b)$.

A similar construction gives the local parametrix $P$ in the neighborhood $U_{\delta}(a)$ of $a$.

\subsection{Final transformation}
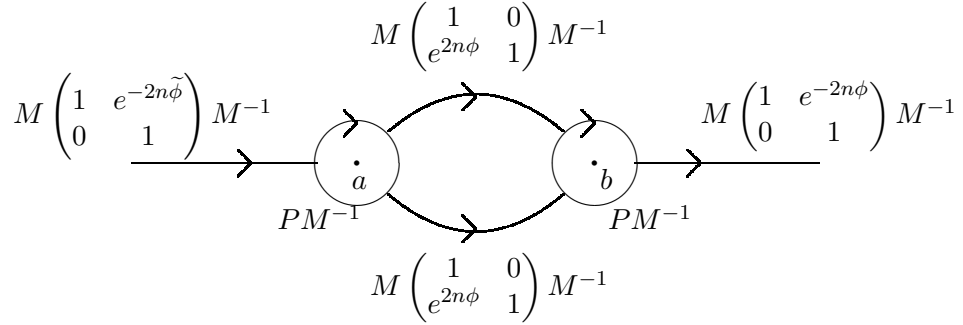
\begin{figure}[t]
\centering
\unitlength 1pt
\linethickness{0.5pt}
\begin{picture}(240,120)(-120,-60)
   \put(55,0){\line(1,0){70}}
   \put(-65,0){\line(-1,0){70}}
   \qbezier(-38,12)(-5,40)(28,12)
   \qbezier(-38,-12)(-5,-40)(28,-12)
   \qbezier(80,0)(79,1)(75,5)   \qbezier(80,0)(79,-1)(75,-5)
   \qbezier(-5,25)(-6,26)(-10,30)   \qbezier(-5,25)(-6,24)(-10,20)
   \qbezier(-5,-25)(-6,-26)(-10,-30)   \qbezier(-5,-25)(-6,-24)(-10,-20)
   \qbezier(-90,0)(-91,1)(-95,5)   \qbezier(-90,0)(-91,-1)(-95,-5)
   \qbezier(-50,15)(-51,16)(-55,20) \qbezier(-50,15)(-51,14)(-55,10)
   \qbezier(40,15)(39,16)(35,20) \qbezier(40,15)(39,14)(35,10)
   \put(-52,-10){$a$}
   \put(-50,0){\circle*{2}}
   \put(-50,0){\circle{30}}
   \put(42,-10){$b$}
   \put(40,0){\circle*{2}}
   \put(40,0){\circle{30}}
   \put(80,15){$M \begin{pmatrix} 1 & e^{-2n\phi}  \\ 0 & 1\end{pmatrix} M^{-1}$}
   \put(-45,45){$M \begin{pmatrix} 1 & 0 \\ e^{2n\phi} & 1 \end{pmatrix} M^{-1}$}
   \put(-45,-50){$M \begin{pmatrix} 1 & 0 \\ e^{2n\phi} & 1 \end{pmatrix} M^{-1}$}
   \put(-180,15){$M\begin{pmatrix} 1 & e^{-2n\widetilde{\phi}} \\ 0 & 1 \end{pmatrix} M^{-1}$}
   \put(45,-25){$PM^{-1}$}
   \put(-80,-25){$PM^{-1}$}
   \end{picture}
   \caption{Contour $\Sigma_R$ and jump matrices for the  RH problem for $R$.}\label{figure4}
\end{figure}

In the final transformation we put
\begin{align} \label{eq:defR}
    R(z) = \begin{cases}
        S(z) M(z)^{-1},  &
        z \in \mathbb C \setminus( \Sigma_S \cup \overline{U_{\delta}(a)} \cup \overline{U_{\delta}(b)}), \\
        S(z) P(z)^{-1}, &
    z \in (U_{\delta}(a) \cup U_{\delta}(b)) \setminus \Sigma_S.
    \end{cases}
    \end{align}
Then $R$ has an analytic continuation to $\mathbb C \setminus \Sigma_R$ where $\Sigma_R$
is shown in Figure~\ref{figure4}.

$R$ solves the following RH problem.
\begin{itemize}
\item[(1)] $R$ is analytic in $\mathbb C \setminus \Sigma_R$.
\item[(2)] $R$ satisfies the jump conditions $R_+ = R_- J_R$
where $J_R$ are the matrices given in Figure \ref{figure4}.
\item[(3)] $R(z) = I + O (z^{-1})$ as $z \to \infty$.
\end{itemize}

The jump matrices tend to the identity matrix as $n \to \infty$. Indeed,
the jump matrix $J_R$ on the boundaries of the disks satisfies
\[ J_R(z) = P(z) M(z)^{-1} = I + O(n^{-1}) \]
as $n \to \infty$, because of the matching condition \eqref{eq:matching}.
On the remaining parts of $\Sigma_R$ we have
that $J_R = I + O(e^{-cn})$ as $n \to \infty$ for some  constant $c > 0$.

Technical estimates on solutions of RH problems, see \cite{Deift1},
now guarantee that in this case
\begin{equation} \label{eq:estimateR}
    R(z) = I + O(1/n) \qquad \text{as } n \to \infty,
    \end{equation}
uniformly for $z \in \mathbb C \setminus \Sigma_R$.

\subsection{Proof of bulk universality} \label{subsec:bulkuniversality}
Now we turn our attention again to the correlation kernel \eqref{eq:kernelinY}
which, since $n = N$, we denote by $K_n$ instead of $K_{n,N}$.
We follow what happens with $K_n$ under the transformations
$Y \mapsto T \mapsto S$. We assume that $x,y \in (a,b)$.

From the definition \eqref{eq:defT} we obtain
\[
    K_n(x,y) = \frac{\sqrt{e^{-nV(x)}} \sqrt{ e^{-nV(y)}}}
    {2\pi i(x-y)}
    \begin{pmatrix} 0 & e^{n(g_+(y)+\ell/2)} \end{pmatrix} T_+^{-1}(y)
    T_+(x) \begin{pmatrix} e^{n(g_+(x) + \ell/2)} \\ 0 \end{pmatrix}
    \]
which by (non-trivial) properties of \eqref{eq:defg} and \eqref{eq:defphi},
based on the variational condition \eqref{eq:varcondition1},
can be rewritten as
\begin{equation} \label{eq:kernelinT}
    K_n(x,y) = \frac{1}
    {2\pi i(x-y)}
    \begin{pmatrix} 0 & e^{-n\phi_+(y)} \end{pmatrix} T_+^{-1}(y)
    T_+(x) \begin{pmatrix} e^{-n\phi_+(x)} \\ 0 \end{pmatrix}.
    \end{equation}
Using the transformation \eqref{eq:defS} in the upper part
of the lens, we see that \eqref{eq:kernelinT} leads to
\begin{equation} \label{eq:kernelinS}
    K_n(x,y) = \frac{1}{2\pi i(x-y)}
    \begin{pmatrix} -e^{n \phi_+(y)} & e^{-n\phi_+(y)} \end{pmatrix}
    S_+^{-1}(y) S_+(x) \begin{pmatrix} e^{-n \phi_+(x)} \\ e^{n\phi_+(x)}
    \end{pmatrix} \end{equation}
for $x, y \in (a,b)$.
This is the basic formula for $K_n$ in terms of $S$.

If $x,y \in (a+\delta, b-\delta)$,
then by \eqref{eq:defR}
\begin{align} \nonumber
    S_+^{-1}(y) S_+(x) = M_+^{-1}(y) R(y)^{-1} R(x) M_+(x).
    \end{align}
The uniform estimate \eqref{eq:estimateR} on $R$ can then be used to show
that
\begin{align} \label{eq:estimateS}
    S_+^{-1}(y) S_+(x) = I + O(x-y).
    \end{align}
    with an $O$-term that is uniform for $x,y \in (a+\delta, b-\delta)$.
Using \eqref{eq:estimateS} in \eqref{eq:kernelinS} we arrive at
\begin{align} \label{eq:kernelinbulk1}
    K_n(x,y) =  \frac{\sin ( i n (\phi_+(y) - \phi_+(x))}{\pi (x-y)}   + O(1).
    \end{align}

Take $x^* \in (a,b)$ fixed and let $c = \rho_V(x^*) > 0$.
We may assume that $\delta > 0$ has been chosen so small
that  $x^* \in (a+\delta, b-\delta)$.
Replace $x$ and $y$ in \eqref{eq:kernelinbulk1}  by $x^* + \frac{x}{cn}$ and $x^* + \frac{y}{cn}$
respectively. Then after dividing through by $c n$, we have for fixed $x$ and $y$,
\begin{multline} \label{eq:kernelinbulk2}
    \frac{1}{c n} K_n\left(x^* + \frac{x}{c n}, x^* + \frac{y}{c n} \right)
    \\
    = \frac{\sin\left(in \left(\phi_+\left(x^* + \frac{y}{c n}\right)
        - \phi_+\left(x^* + \frac{x}{cn}\right)\right)\right)}{\pi (x-y)}
    + O\left(\frac{1}{n} \right)
    \end{multline}
as $n \to \infty$, and the $O$-term is uniform for $x$ and $y$ in compact
subsets of $\mathbb R$.
Since $\phi_+'(s) = \pi i \rho_V(s)$ for $s \in (a,b)$, and $c = \rho_V(x^*)$,
we have that
\begin{align*}
    in \left(\phi_+\left(x^* + \frac{y}{c n}\right)
        - \phi_+\left(x^* + \frac{x}{c n}\right) \right)
            = \pi(x-y) + O\left(\frac{x-y}{n}\right)
        \end{align*}
and therefore the rescaled kernel \eqref{eq:kernelinbulk2}
does indeed tend to the sine kernel \eqref{eq:sinekernel}
as $n \to \infty$. This proves the bulk universality.

\paragraph{Remark.}
Letting $y \to x$ in \eqref{eq:kernelinbulk1} and dividing by $n$
we also obtain
    \begin{align*}
        \frac{1}{n} K_n(x,x)
            & = \frac{1}{\pi i} \phi_+'(x) + O \left(\frac{1}{n} \right) \\
            & = \frac{1}{\pi} h(x) \sqrt{(b-x)(x-a)} + O \left(\frac{1}{n} \right)
                && \text{(by \eqref{eq:defphi})} \\
            & = \rho_V(x) + O \left(\frac{1}{n} \right) && \text{(by \eqref{eq:rhoV})}
            \end{align*}
as $n \to \infty$, uniformly for $x$ in compact subsets of $(a,b)$,
which proves that $\rho_V$ is the limiting means eigenvalue density.

\subsection{Proof of edge universality}

We will not give the proof of the edge universality in detail.
The proof starts from the representation \eqref{eq:kernelinS}
of the correlation kernel in terms of $S$.
In the neighborhood of $a$ and $b$ we have by \eqref{eq:defR} that $S = RP$,
where $P$ is the local parametrix that is constructed out of Airy functions,
as it involves the solution  of the Airy Riemann-Hilbert problem.

From \eqref{eq:defA}  it can be checked
that the Airy kernel \eqref{eq:Airykernel} is given in terms of
the solution $A(z)$ by
\[
    K^{\Ai}(x,y)  =
\frac{1}{2\pi i(x-y)}
    \times \begin{cases}
    \begin{pmatrix} 0 & 1 \end{pmatrix}
    A_+^{-1}(y) A_+(x) \begin{pmatrix} 1 \\ 0 \end{pmatrix}
    & \text{if both $x,y > 0$}, \\
    \begin{pmatrix} -1 & 1 \end{pmatrix}
    A_+^{-1}(y) A_+(x) \begin{pmatrix} 1 \\ 0 \end{pmatrix}
    & \text{if $x> 0$ and $y < 0$}, \\
    \begin{pmatrix} 0 & 1 \end{pmatrix}
    A_+^{-1}(y) A_+(x) \begin{pmatrix} 1 \\ 1 \end{pmatrix}
    & \text{if $x < 0$ and $y > 0$}, \\
 \begin{pmatrix} -1 & 1 \end{pmatrix}
    A_+^{-1}(y) A_+(x) \begin{pmatrix} 1 \\ 1 \end{pmatrix}
    & \text{if both $x,y < 0$},
    \end{cases} \]
and this is exactly what comes out of the calculations for the
scaling limit of the eigenvalue correlation kernels $K_n$ near
the edge point.

\sect{Non-standard universality classes} \label{sect:non-standard}

The standard universality classes (sine, Airy and Bessel)
describe  the local eigenvalue statistics around regular points.

In the unitary ensemble \eqref{eq:unitaryensemble}
there are three types of singular eigenvalue behavior.
They all depend on the behavior of the global eigenvalue
density $\rho_V$.
The three types of singular behavior are:
\begin{itemize}
\item The density $\rho_V$ vanishes at an interior point of the support.
\item The density $\rho_V$ vanishes to higher order at an edge point of the support (higher than square root).
\item Equality holds in the variational inequality \eqref{eq:varcondition2} at a point
outside $\supp(\mu_V)$.
\end{itemize}

In each of these cases there exists a family of limiting correlation kernels
that arise in a double scaling limit. In \eqref{eq:unitaryensemble}
and \eqref{eq:pdfunitary} one lets $n, N \to \infty$ with $n/N \to 1$
at a critical rate so that for some exponent $\gamma$, the limit
\begin{equation} \label{eq:limitnN}
    \lim_{n \to \infty} n^{\gamma} \left(\frac{n}{N} - 1 \right)
    \end{equation}
exists.  The family of limiting kernels is  parametrized
by the value of the limit \eqref{eq:limitnN}.

Most rigorous results that have been obtained in this direction are based
on the RH problem, and use an extension of the steepest descent analysis
that was discussed in the previous section. Non-standard universality classes
are also discussed in Chapters 12 and 13.

\subsection{Interior singular point}
An interior singular point is a point $x^*$ where $\rho_V$ vanishes.
Varying the parameters in $V$ may then either lead to a gap in the
support around $x^*$, or to the closing of the gap.
The local scaling limits at $x^*$ depend on the order of vanishing at $x^*$,
that is, on the positive integer $k$ so that
\[ \rho_V(x) = c(x - x^*)^{2k} (1 + o(1)) \qquad \text{as } x \to x^* \]
with $c > 0$.

The case of quadratic vanishing (i.e., $k=1$) was considered in \cite{BI2}
for the critical quartic potential
\[ V(x) = \frac{1}{4} x^4 - x^2 \]
and in \cite{CK, Sh1} for more general $V$. For $k=1$ one takes
$\gamma = 2/3$ in \eqref{eq:limitnN}. The limiting kernels are parametrized
by a parameter $s$ which is proportional to the limit \eqref{eq:limitnN}
\[ s =  c  \lim_{n \to \infty} n^{2/3} \left( \frac{n}{N} - 1\right), \]
where the proportionality constant $c > 0$ is (for the case $\supp(\mu_V) = [a,b]$)
\[ c = \frac{2}{(\pi \rho_V''(x^*))^{1/3}  \sqrt{(b-x^*)(x^*-a)}}. \]
The $s$-dependence is then governed by the Hastings-Mcleod
solution $q(s)$ of the Painlev\'e II equation
\begin{equation} \label{eq:PainleveII}
    q'' = sq + 2q^3.
    \end{equation}
The limiting kernels are therefore called Painlev\'e II kernels
and we denote them by $K^{\PII}(x,y;s)$, even though $q(s)$ itself
does not appear in the formulas for the kernels. What does appear is a solution
of the Lax pair equations
\begin{equation} \label{eq:Laxpair}
\begin{aligned}
    \frac{\partial}{\partial x} \begin{pmatrix} \Phi_1(x;s) \\ \Phi_2(x;s) \end{pmatrix}
    & = \begin{pmatrix} -4 i x^2 - i(s+ 2q^2) & 4xq + 2ir \\ 4xq - 2ir & 4i x^2 + i(s+2q^2) \end{pmatrix}
    \begin{pmatrix} \Phi_1(x;s) \\ \Phi_2(x;s) \end{pmatrix}
    \\
    \frac{\partial}{\partial s} \begin{pmatrix} \Phi_1(x;s) \\ \Phi_2(x;s) \end{pmatrix}
    & = \begin{pmatrix} - i x & q \\ q & ix \end{pmatrix}
    \begin{pmatrix} \Phi_1(x;s) \\ \Phi_2(x;s) \end{pmatrix}
\end{aligned} \end{equation}
where $q = q(s)$ is the Hastings-Mcleod solution of \eqref{eq:PainleveII}
and $r = r(s) = q'(s)$.

There is a specific solution of \eqref{eq:Laxpair}
so that the family of limiting kernels takes the form
\begin{equation} \label{eq:PIIkernel}
    K^{\PII}(x,y;s) =  \frac{-\Phi_1(x;s) \Phi_2(y;s) + \Phi_1(y; s)\Phi_1(y; s)}{2\pi i(x-y)}.
    \end{equation}

For $k \geq 2$, one takes $\gamma = 2k/(2k+1)$ in \eqref{eq:limitnN},
and  the limiting kernels can be described in a similar way by
a special solution of the $k$th member of the Painlev\'e II hierarchy.
The functions in the kernels itself are solutions of the associated Lax pair equations.

The connection with Painlev\'e II also holds for
the unitary matrix model with a spectral singularity \eqref{eq:spectralsingularitymodel}
\[   \frac{1}{\tilde{Z}_{n,N}} \left| \det M \right|^{2\alpha}
     e^{-N \Tr V(M)} dM, \qquad \alpha > -1/2. \]
In the multicritical case where $\rho_V$ vanishes quadratically at $x=0$,
the limiting kernels are associated with a special solution
of the general form of the Painlev\'e II equation
\[ q'' = sq + 2 q^3 - \alpha \]
with parameter $\alpha$, see \cite{ADMN2,CKV}.

\subsection{Singular edge point}
A singular edge point is a point $x^*$ where $\rho_V$ vanishes to higher
order than square root. The local scaling limits depend again on the order of vanishing.
If $x^*$ is a right-edge point of an interval
in the support, then there is an even integer $k$ such that
\begin{equation} \label{eq:singularedgedensity}
    \rho_V(x) = c (x^*-x)^{k+1/2}(1+o(1)) \qquad \text{as } x \nearrow x^*
    \end{equation}
with $c > 0$. Here one takes $\gamma = (2k+2)/(2k+3)$ in \eqref{eq:limitnN}.
The limiting kernels are described by Lax pair solutions associated
with a special solution of
the $k$th member of the Painlev\'e I hierarchy. For $k=2$ this
is worked out in detail in \cite{CV}. For general $k$, and assuming $x^*$
is the right-most point in the support, the largest eigenvalue
distributions are studied in \cite{CIK}.
These generalizations of the Tracy-Widom distribution are expressed
in terms of members of a Painlev\'e II hierarchy.

The case $k = 1$ in \eqref{eq:singularedgedensity} cannot occur
in the unitary matrix model \eqref{eq:unitaryensemble}, although
it frequently appears in the physics literature, see e.g.\  \cite{DGZ},
where, for example, it is  associated with the potential
\[ V(x) = - \frac{1}{48} x^4 + \frac{1}{2} x^2. \]
This potential does not satisfy \eqref{eq:growthV} and the
model  \eqref{eq:unitaryensemble} is not well-defined as a
probability measure on Hermitian matrices, although it can be studied
in a formal sense, see e.g.\ in Chapter 16.

In \cite{FIK} and \cite{DK} the polynomials that are orthogonal on
certain contours in the complex plane with a weight $e^{-NV(x)}$,
where $V(x) = t \frac{x^4}{4} + \frac{x^2}{2}$ and $t \approx -\frac{1}{12}$,
are studied and the connection with special solutions of the Painlev\'e~I equation
\[ q'' = 6q^2 + s \]
is rigorously established.

\subsection{Exterior singular point}
An exterior point $x^*$ is singular if  there is equality in the
variational inequality \eqref{eq:varcondition2}. At such a point
a new interval in the support may arise when perturbing the potential $V$.
The limiting kernels at such a point depend on the order of vanishing
of
\[ 2 \int \log \frac{1}{|x-y|} d\mu_V(y) + V(x) - \ell \]
at $x = x^*$.

In this situation the appropriate scaling is so that
\begin{equation} \label{eq:limitnN2}
    \lim_{n \to \infty} \frac{n}{\log n} \left(\frac{n}{N} - 1 \right)
    \end{equation}
exists, instead of \eqref{eq:limitnN}. This special kind of scaling was
discussed in \cite{Ake, Eyn}.
Maybe surprisingly, there is no connection with Painlev\'e equations
in this case. In the simplest case of
quadratic vanishing at $x=x^*$ the possible limiting kernels
are finite size GUE kernels and certain interpolants, see \cite{BL,Claeys,Mo}.

\subsection{Pearcey kernels}
The Painlev\'e II kernels \eqref{eq:PIIkernel} are the canonical kernels
that arise at the closing of a gap in  unitary ensembles.

Br\'ezin and Hikami \cite{BH} were the first to identify a second
one parameter family of kernels that may arise at the closing of a gap.
This is the family of Pearcey kernels
\begin{equation} \label{eq:Pearceykernel}
    K^{\Pear} (x,y; s) = \frac{p(x) q''(y) - p'(x) q'(y) + p''(x) q(y) - s p(x) q(y)}{x-y}
    \end{equation}
with $s \in \mathbb R$, where $p$ and $q$ are solutions of the Pearcey differential equations
$p'''(x) = x p(x) - s p'(x)$ and $q'''(y) = yq(y) + s q'(y)$.
The kernel is also given by the double integral
\begin{equation} \label{eq:Pearceydoubleintegral}
    K^{\Pear}(x,y;s) =
    \frac{1}{(2\pi i)^2} \int_C \int_{-i\infty}^{i\infty}
        e^{- \frac{1}{4} \eta^4 + \frac{s}{2} \eta^2 - \eta y + \frac{1}{4} \xi^4 - \frac{s}{2} \xi^2 + \xi x}
            \frac{d\eta \, d\xi}{\eta-\xi}
            \end{equation}
where the contour $C$ consists of the two rays from $\pm \infty e^{i\pi/4}$ to $0$
together with the two rays from $0$ to $\pm \infty e^{-i \pi/4}$.

The Pearcey kernels appear at the closing of the gap
in the Hermitian matrix model with external source
\begin{equation} \label{eq:sourcemodel}
    \frac{1}{\tilde{Z}_{n}} e^{- n \Tr( V(M) - AM)} d M
    \end{equation}
where the external source $A$ is a given Hermitian $n \times n$ matrix.
This was proved in \cite{BH} for the case where $V(x) = \frac{1}{2} x^2$
and $A$ is a diagonal matrix
\begin{equation} \label{eq:externalsourceA}
     A = \diag(\underbrace{a, \ldots, a}_{n/2 \text{ times}},
        \underbrace{-a, \ldots, -a}_{n/2 \text{ times}})
        \end{equation}
depending on the parameter $a > 0$. For $a > 1$ the eigenvalues of $M$ accumulate
on two intervals as $n \to \infty$ and for $0 < a < 1$ on one interval.
The Pearcey kernels \eqref{eq:Pearceydoubleintegral} appear in a double scaling
limit around the critical value $a=1$, see also \cite{BK} for an analysis
of an  associated $3\times 3$ matrix valued
RH problem. The matrix model with external source \eqref{eq:sourcemodel}
with quadratic potential has an interesting interpretation in terms of non-intersecting
Brownian motions \cite{ABK}.

The case of a quartic polynomial potential $V(x) = \frac{1}{4} x^4 - \frac{t}{2} x^2$
was analyzed recently in \cite{BDK}. Here it was found that the closing of
the gap can be either of the Pearcey type or of the Painlev\'e II type, depending
on the value of $t \in \mathbb R$.

\end{document}